\documentstyle[12pt,amssymb]{article}
\def\be{\begin{equation}}
\def\bea{\begin{eqnarray}}     
\def\ee{\end{equation}}
\def\eea{\end{eqnarray}}
\newcommand{\la}{\label}
\def\M{{\bf M}^{n+1}}

\def\ad{{\rm ad}_{{\cal L}, {{\cal L}_{0}}}^{M}}
\def\eq#1{(\ref{#1})}

\begin{document}
\begin{center}
{\Large \bf 
Huygens' Principle in Minkowski Spaces and Soliton\\*[1ex]
Solutions of the Korteweg-de Vries Equation}\\*[3ex] 
{\large \bf Yuri~Yu.~Berest} and {\large \bf Igor~M.~Loutsenko}\\*[1ex] {\it
Universit\'e de Montr\'eal, Centre de Recherches Math\'ematiques,\\
C.P.~6128, succ. Centre-Ville, Montreal (Quebec), H3C 3J7, Canada,\\
e-mail: beresty@crm.umontreal.ca\ ,\ loutseni@crm.umontreal.ca}
\end{center}
\vspace{1.5cm}
\begin{center}
\bf Abstract
\end{center}
\begin{quote}
 A new class of linear second order hyperbolic partial differential operators
satisfying Huygens' principle in Minkowski spaces is presented. The 
construction reveals a direct connection between Huygens' principle and
the theory of solitary wave solutions of the Korteweg-de Vries 
equation.
 \end{quote}

\vspace{1cm}


\newpage

\section*{I. Introduction}

  The present paper deals with the problem of describing all linear 
second order
partial differential operators for which Huygens' principle is valid in the sense
of ``Hadamard's minor premise''. Originally posed by J.Hadamard in his Yale
lectures on hyperbolic equations \cite{Had}, this problem
is still far from being completely solved\footnote {Hadamard's problem, 
or {\it the problem of diffusion of waves}, has received a good deal of attention
and the literature is extensive (see, e.g., \cite{BV}, \cite{McL1}, \cite{Cou}, \cite{Fri},
\cite{PG}, \cite{Gun}, \cite{Hel}, \cite{Ibr}, \cite{McL},  and references 
therein). For a historical account we refer the reader to the 
articles \cite{Dui}, \cite{Gun1}.}.

  The simplest examples of Huygens' operators are the ordinary wave operators
\begin{equation}
\label{1}
\Box_{n+1} = \left( \frac{\partial}{\partial x^{0}}\right)^{2} - \left( \frac{\partial}{\partial x^{1}}\right)^{2} - 
 \ldots - \left( \frac{\partial}{\partial x^{n}}\right)^{2} 
\end{equation}
in an odd number $ \, n \geq 3\, $ of space dimensions and those ones reduced to
(\ref{1}) by means of elementary transformations, i.e. by local nondegenerate
 changes of
coordinates $\  x \mapsto f(x)\ $; gauge and conformal transformations of a
given operator $ {\cal L} \mapsto \theta(x) \circ {\cal L } \circ  \theta(x)^{-1}\ , \  {\cal L} \mapsto 
\mu (x) {\cal L} $ with some locally smooth nonzero functions $ \theta(x) $ and
$ \mu (x) $. These operators are usually called {\it trivial Huygens' operators},
and the famous ``Hadamard's conjecture'' claims that all Huygens' operators 
are trivial.

  Such a strong assertion turns out to be valid only for (real) Huygens' operators with a
constant principal symbol in $\, n=3\, $ \cite{Mat}. Stellmacher \cite{Stell} found the first
non-trivial examples of hyperbolic wave-type operators satisfying Huygens' principle, 
and thereby disproved Hadamard's conjecture in higher dimensional Minkowski spaces. 
Later Lagnese \& Stellmacher \cite{Stell1} extended these examples and even solved \cite{Lag}
Hadamard's problem for a restricted class of hyperbolic operators, namely
\begin{equation}
\label{2}
{\cal L}  = \Box_{n+1} + u(x^0)\ ,
\end{equation}
where $\, u\left( x^0 \right)\, $  is an analytic function (in its domain of definition)
depending on a single variable only. 
It turns out that the potentials $ u(z) $ entering into (\ref{2})  are rational functions which can be expressed
explicitly in terms of some polynomials\footnote{This remarkable class of polynomials
seems to have been found for the first time by Burchnall and Chaundy \cite{BCh}.} $ {\cal P}_k \left( z \right)$:
\begin{equation}
\label{3}
u(z) = 2 \left( \frac{d}{dz}\right)^2 \log {\cal P}_k \left( z \right)\ , \quad k=0,1,2, \ldots ,
\end{equation}
the latter being defined via the following differential-recurrence relation:
\begin{equation}
\label{4}
{\cal P}_{k+1}' {\cal P}_{k-1}- {\cal P}_{k-1}' {\cal P}_{k+1} =
(2 k+1) {\cal P}_{k}^{2}\ , \quad {\cal P}_{0} = 1\ , \ {\cal P}_{1} = z\ .
\end{equation}
Since the works of Moser et al.  \cite{AM}, \cite{AMM} the 
potentials (\ref{3}) are known as
{\it rational solutions} of the Korteweg-de Vries equation decreasing at infinity\footnote{ The
coincidence of such rational solutions of the KdV-hierarchy with the Lagnese-Stellmacher
potentials has been observed by Schimming \cite{Sch1}, \cite{Sch2}. }. 

\par   A wide class of Huygens' operators in Minkowski spaces has been
discovered recently by Veselov and one of the authors \cite{BV1}, \cite{BV2} 
(see also the review article \cite{BV}). 
These operators can also be presented in a self-adjoint form
\begin{equation}
\label{5}
{\cal L}  = \Box_{n+1} + u(x)
\end{equation}
with a locally analytic potential $ u \left( x \right) $ depending on several
variables. More precisely, $ u \left( x \right) $ belongs to the class of so-called
{\it Calogero-Moser potentials} associated with finite reflection groups ({\it
Coxeter groups}):
\begin{equation}
\label{6}
u(x)= \sum\limits_{\alpha \in \Re_{+}}^{} \frac{m_{\alpha}(m_{\alpha} + 1) (\alpha, \alpha)}
{(\alpha, x)^2}\ .
\end{equation}
In formula (\ref{6}) $ \Re_{+} \equiv \Re_{+} ({\cal G}) $ stands for a properly chosen and 
oriented subset of normals to reflection hyperplanes of a Coxeter group $ {\cal G}
$. The group $ {\cal G} $  acts on $ {\bf M} ^ {n+1} $ in such a way that the time
direction is preserved. The set $ \left\{ m_{\alpha} \right\} $  is a  collection of
non-negative integer labels attached to the normals $ {\alpha} \in \Re $ so that $
m_{w(\alpha )} = m_{ \alpha } $  for all $ w \in {\cal G} . $ 
Huygens'     principle holds for (\ref{5}), (\ref{6}), provided $ n $ is odd, and
\begin{equation}
\label{7}
n \geq 3 + 2 \sum\limits_{\alpha \in \Re_{+}}^{} m_{\alpha}\ .
\end{equation} 

In the present work we construct a new class of self-adjoint wave-type
operators (\ref{5}) satisfying  Huygens'     principle in Minkowski spaces. As we will see, this class
provides a natural extension of the hierarchy of Huygens' operators associated to
Coxeter groups. On the other hand, it turns out to be related in a surprisingly simple and 
fundamental way to the theory of solitons.

To present the construction we consider a $(n+1)$-dimensional Minkowski
space $ { \bf M }^{ n+1 }  \cong  { \bf R }^{ 1 , n } $ with the metric signature 
$ ( + , -, -, \ldots, -) $ and fixed time direction $ \theta \in { \bf M }^{ n+1 } $. 
We write $ {\bf Gr}_{\perp} (n+1,2) \subset {\bf Gr}(n+1,2) $  for a set 
of all 2-dimensional space-like linear subspaces in $ { \bf M }^{n+1} $ 
orthogonal to $ \theta $.
Every 2-plane $ E \in {\bf Gr}_{  \perp  } (n+1,2) $ is equipped with the usual Euclidean
structure induced from  $ {\bf M }^{ n+1 } $.   To define the potential $ u(x) $
we fix such a plane $ E $ and introduce polar coordinates $ ( r, \varphi ) $ therein.

  Let  $ (k_{i})^{N}_{i=1} $ be a strictly increasing sequence of integer positive
num\-bers\footnote{Using the terminology adopted in the group representation theory we will 
call such integer monotonic sequences {\it partitions}.}: 
$ 0 \leq k_1 < k_2 < \ldots < k_{N-1} < k_N $, 
and let $ \left\{  \Psi_{i} ( \varphi ) \right\} $
be a set of $ 2 \pi$-periodic functions on $ {\bf R}^{1}$:
\begin{equation}
\label{8}
\Psi_{i}(\varphi) :=  \cos (k_{i} \,\varphi + \varphi_{i})\ , \quad  \varphi_{i} \in {\bf R}\ ,
\end{equation} 
associated to $ (k_{i}) $.  The Wronskian of this set
\begin{equation}
\label{9}
{\cal W}\left[\Psi_{1}, \Psi_{2}, \ldots , \Psi_{N} \right] := \det 
\left(
\begin{array}{cccc}
\Psi_{1}(\varphi) & \Psi_{2} (\varphi)  & \ldots & \Psi_{N} (\varphi)\\
\Psi_{1}'(\varphi) & \Psi_{2}'(\varphi) & \ldots & \Psi_{N}'(\varphi) \\
\vdots  &  \vdots  & \ddots & \vdots \\
\Psi_{1}^{(N-1)} (\varphi) & \Psi_{2}^{(N-1)}(\varphi) & \ldots &
 \Psi_{N}^{(N-1)}(\varphi) 
\end{array}
\right)
\end{equation} 
does not vanish indentically since  $ \Psi_{i}(\varphi) $ are linearly independent. 

Let 
$$
 \Xi := \left\{  x \in \M \ | \  r^{|k|}\, {\cal W}\left[\Psi_{1}, \Psi_{2}, \ldots , \Psi_{N} \right] = 0 \ ,
 \ |k| := \sum_{i=1}^{N} k_{i} \right\} 
$$ 
be an algebraic hypersurface of zeros of the Wronskian \eq{9}
in the Minkowski space $ \M $, and let $ \Omega \subset \M \setminus \Xi $
be an open connected part in its complement. 

We define $ u(x) $ in terms of cylindrical coordinates in $ \M $ with polar 
components in $ E $~:
\be
\la{10}
u = u_{k}(x) := - \frac{2}{r^2} \left( \frac{\partial}{ \partial \varphi}\right)^2 \log 
{\cal W}\left[\Psi_{1}(\varphi), \Psi_{2}(\varphi), \ldots , \Psi_{N}(\varphi) \right] \ .
\ee
It is easy to see that in a standard Minkowskian coordinate chart $ u(x) $ is a real {\it rational}
function on $ \M $ having its singularities on $ \Xi $. In particular, it is locally analytic
in $ \Omega $.

Our main result reads as follows.

\vspace{0.3cm}

{\bf Theorem.}\ {\it Let $\ \M \cong {\bf R}^{1,n} $ be a Minkowski space, and let }
\be
\la{11}
{\cal L}_{(k)} := \Box_{n+1} + u_{k}(x)
\ee
{\it be a wave-type second order hyperbolic operator with the potential \eq{10} associated to
an arbitrary strictly monotonic partition $ (k_i) $ of height $ N $~: 
$$
 0 \leq k_1< k_2 <  \ldots < k_{N}\ , \quad  k_{i} \in {\bf Z}\ ,\quad  i=1,2, \ldots, N\ . 
$$ 
Then operator $ {\cal L}_{(k)} $ satisfies
Huygens'   principle at every point $ \xi \in \Omega $, provided $ n $ is odd, and
}
\be
\la{12}
n \geq 2\, k_{N} + 3\ .
\ee

\vspace{0.3cm}

{\it Remark I.} \  A similar result is also valid if one takes an arbitrary Lorentzian  2-plane $ H \in {\bf Gr}_{\parallel}(n+1,2) $
in the Minkowski space $ \M $ containing the time-like vector $ \theta $. More precisely, in this case the potential $ u_{k}(x) $
associated to the partition $ (k_{i}) $ is introduced in terms of pseudo-polar coordinates $ (\varrho,\vartheta) $ in  $ H $:
\be
\la{13}
u_{k}(x) := - \frac{2}{{\varrho}^2} \left( \frac{\partial}{ \partial \vartheta}\right)^2 \log 
{\cal W}\left[\psi_{1}, \psi_{2}, \ldots , \psi_{N} \right] \ ,
\ee
where $ x^{0} = \varrho \sinh \vartheta\,$, and, say, $\, x^{1} = \varrho \cosh \vartheta\, $.
The functions $ \psi_{i} $ involved in \eq{13} are given by
\be
\la{14}
\psi_{i} =  \cosh (k_{i}\, \vartheta + \vartheta_{i})\ , \quad \vartheta_{i} \in {\bf R}\  .
\ee
The theorem formulated above holds when the potential \eq{10} is replaced by \eq{13}.

\vspace{0.2cm}

{\it Remark II.} \ The potentials \eq{10} considerably extend  the class of Calogero-Moser potentials \eq{6} related to
 Coxeter groups of rank $ 2 $. Indeed, in $ {\bf R}^{2} $ any Coxeter group $ {\cal G}  $ is a dihedral group $ I_{2}(q) $,
i.e. the group of symmetries of a regular $2q$-polygon. It has one or two conjugacy classes of reflections according 
as $ q $ is odd or even. The corresponding potential \eq{6} can be rewritten in terms of polar coordinates as follows
(see  \cite{OP}):
$$
u(r,\varphi) = \frac{m(m+1)q^2}{r^2 \sin^{2}(q\,\varphi)} \ ,\quad \mbox{\rm when}\ q \ \mbox{\rm odd} \ ,
$$
and
$$
u(r,\varphi) = \frac{m(m+1)\left(q/2\right)^2}{r^2 \sin^{2} (q/2)\,\varphi}  + 
                     \frac{m_{1}(m_{1}+1) \left(q/2 \right)^2}{r^2 \cos^{2} (q/2)\,\varphi} 
\ , \quad \mbox{\rm when}\ q \ \mbox{\rm even}\ . 
$$
It is easy to verify that  formula \eq{10} boils down to these  forms  if we fix $\, N := m \, ; \, 
 \varphi_{i} := (-1)^{i} \pi /2 \, ,\, i =1,2, \ldots, N\, $, and
choose  
$$
 k := (q, \, 2q, \, 3q, \, \ldots ,\, mq ) \ , 
$$ 
when $ q $ is odd, and   
$$
 k  := (\frac{q}{2} , \, q, \, \frac{3q}{2}, \, \ldots ,\, (m-m_1)\frac{q}{2} , \, q+  (m-m_1)\frac{q}{2} , \, 
 2q+  (m-m_1)\frac{q}{2}, \, \ldots\, ,\, (m+m_1)\frac{q}{2} )\ , 
$$
when  $ q $ is even and $\, m > m_{1} \,$, respectively.

\vspace{0.2cm}

{\it Remark III.}\ Let us set $ \varphi_{i} = 4 k_{i}^{3} t + \varphi_{0i} $ and $ \vartheta_{i} = - 4 k_{i}^{3} t + \vartheta_{0i} \ ,
\ i=1,2, \ldots, N \ ;\ \varphi_{0i}  ,  \vartheta_{0i} \in {\bf R}. $ The angular parts of potentials \eq{10}, \eq{13}, i.e.
\be
\la{15}
v(\varphi) = -2 \left( \frac{ \partial }{ \partial \varphi } \right )^2 \log {\cal W} \left[\Psi_{1}(\varphi) , \Psi_{2}(\varphi) , \ldots , 
\Psi_{N}(\varphi) \right] \ , 
\ee
\be
\la{16}
v(\vartheta) = -2 \left( \frac{ \partial }{ \partial \vartheta } \right )^2 \log {\cal W} \left[\psi_{1}(\vartheta) , \psi_{2}(\vartheta) , \ldots , 
\psi_{N}(\vartheta) \right] \ , 
\ee
are known (see, e.g., \cite{Deift}, \cite{MS}) to be respectively singular {\it periodic}  and proper {\it N-soliton} solutions of 
the Korteweg-de Vries equation
\be
\la{17}
v_{t} = - v_{\varphi \varphi \varphi} + 6 v v_{\varphi}\ .
\ee
It is also well-known that $N$-soliton potentials \eq{16} constitute the  whole class  of so-called {\it reflectionless}
real potentials for the  
one-dimensional Schr\"odinger operator $ L = -  \partial^2 / \partial \vartheta^2 + v(\vartheta) $
(see, e.g., \cite{AC}).

\vspace{0.2cm}

In conclusion of this section we put forward the following 
conjecture\footnote{
{\bf Note added in the proof.}\ This conjecture has been proved recently by one of the
authors in \cite{Ber11}.
}. 

\vspace{0.3cm}

{\bf Conjecture.}\ 
The wave-type operators \eq{11} with potentials of the form \eq{10} give {\it a complete} solution
of Hadamard's problem in Minkowski spaces $ \M $ within a restricted class of linear second order hyperbolic operators
$$
{\cal L} = \left( \frac{\partial}{\partial x^{0}}\right)^{2} - \left( \frac{\partial}{\partial x^{1}}\right)^{2}  
- \left( \frac{\partial}{\partial x^{2}}\right)^{2} -
\ldots - \left( \frac{\partial}{\partial x^{n}}\right)^{2} + u(x^1, x^2) 
$$
with real locally analytic potentials $ u = u(x^1, x^2) $
depending on {\it two} spatial variables and homogeneous of degree $ (- 2) $: 
$\, u(\alpha x^1, \alpha x^2) = \alpha^{-2} u(x^1, x^2)\,,\, \alpha >0\,$.

\vspace{0.4cm}

{ \small
{\bf Acknowledgements.}\ 
One of the authors (Yu.~B.) is grateful to Prof. A.~Veselov 
(Loughborough University, UK) and
his collaborators Dr.~O.~Chalykh and M.~Feigin who have 
kindly informed him \cite{Ves} about their recent results 
showing the existence of new algebraically 
integrable Schr\"odinger  operators which are not related to Coxeter root systems. In fact, 
this observation was a motivation for the present work.
We would like also to thank Prof.~P.~Winternitz  (CRM, Universit\'e de Montr\'eal)
for his encouragement and highly stimulating discussions.

The work of Yu. B. was partially supported by the fellowship from Institut
des Sciences Math\'ematiques (Montr\'eal) which is gratefully acknowledged.
The second author (I.~L.) is grateful to Prof.~L.~Vinet  for his support.

}

\section*{II. Huygens'      principle and Hadamard-Riesz expansions}

The proof of the  theorem stated above rests heavily on the Hadamard theory of Cauchy's problem
for  linear second order hyperbolic partial differential equations. Here, we summarize briefly some
necessary results from this theory following essentially M.~Riesz's approach \cite{Riesz} (see also 
\cite{Fri}, \cite{Gun}).

Let $ \M \cong {\bf R}^{1,n} $ be a Minkowski space, and let $ \Omega $ be an open connected part
in $ \M $. We consider a (formally) self-adjoint scalar wave-type operator
\be
\la{18}
{\cal L} = \Box_{n+1} + u(x)\ ,
\ee
defined in $ \Omega $, the scalar field  (potential) $ u(x) $ being assumed to be in
$ {\cal C}^{\infty}(\Omega) $. For any $ \xi \in \Omega $, we define a cone of isotropic (null) vectors
in $ \M $ with its vertex at $ \xi $: 
\be
\la{19}
\gamma(x,\xi) := (x^0- \xi^0)^{2} - (x^1- \xi^1)^{2} - 
\ldots - (x^{n} - \xi^{n})^{2} = 0\ ,
\ee
and single out the following sets~:
\be
\la{20}
\begin{array}{lcr}
C_{\pm}(\xi) & := & \left\{ x \in \M\  | \ \gamma(x,\xi) = 0\,,\, 
\xi^0 \lessgtr x^{0} \right\}\ ,\\*[2ex] 
J_{\pm}(\xi) & := & \left\{ x \in \M\ | \ \gamma(x,\xi) > 0\,,\,
\xi^0 \lessgtr x^{0} \right\}\ .
\end{array}
\ee

\vspace{0.3cm}

{\bf Definition.}\ {\it A (forward) Riesz kernel} of operator $ {\cal L} $ is a holomorphic (entire analytic)
mapping  $ \lambda \mapsto  \Phi_{\lambda}^{\Omega}(x,\xi)\,,\,
\lambda \in \mbox{\bf C} $,
with values in the space of distributions\footnote{By a distribution $ f \in {\cal D}'(\Omega) $ 
we mean, as usual, a linear continuous form on the space $ {\cal D}(\Omega) $ of ${\cal C}^{\infty}$-functions
with supports compactly imbedded in $ \Omega $ (cf., e.g., \cite{GSh}).} 
$ {\cal D}'(\Omega) $, such that for any $ \xi \in \Omega $:
\be
\la{21}
\begin{array}{lcl}
& (i) &\ \mbox{\rm supp}\, \Phi_{\lambda}^{\Omega}(x,\xi) \subseteq 
        \overline{J_{+}(\xi)}\ ,\\*[2ex]
& (ii) &\ {\cal L}\left[ \Phi_{\lambda}^{\Omega}(x,\xi)\right] = 
         \Phi_{\lambda-1}^{\Omega}(x,\xi)\ ,\\*[2ex] 
& (iii) &\ \Phi_{0}^{\Omega}(x,\xi) = \delta (x -\xi)\ .
\end{array}
\ee

\vspace{0.3cm}

The value of the Riesz kernel $\  \Phi_{1}^{\Omega}(x,\xi) := \Phi_{+}(x,\xi)\ $ at $ \lambda = 1 $ 
is  called {\it a (forward) fundamental solution} of the operator $ {\cal L} $:
\begin{equation}
\label{22}
{\cal L} [\Phi_{+}(x, \xi)] = \delta(x-\xi) \ , \quad  \mbox{supp}\ \Phi_{+}(x, \xi) \subseteq \overline{J_{+}(\xi)} \ . 
\end{equation}

Such a solution is known to exist for any $ u(x) \in  {\cal C}^{\infty}(\Omega) $, and it is uniquely determined.

\vspace{0.3cm}

{\bf Definition.} \  The operator  $ {\cal L} $  defined by \eq{18} satisfies {\it Huygens'     principle} in a domain 
$ \Omega_{0} \subseteq \Omega $ in $ \M $ if
\begin{equation}
\label{23}
\mbox{\rm supp} \ \Phi_{+}(x, \xi) \subseteq \overline{C_{+}(\xi)} = \partial J_{+}(\xi) \ .		
\end{equation}
for every point $ \xi \in \Omega_{0} $.

\vspace{0.3cm}

The analytic description of singularities of  Riesz kernel distributions (and, in particular, fundamental solutions)
for second order hyperbolic  differential operators  is given in terms of their asymptotic expansions in the vicinity
of the characteristic cone by a graded scale of distributions with weaker and weaker singularities. Such
``asymptotics in smoothness'', usually called {\it Hadamard-Riesz expansions}, turn out to be  very important
for testing Huygens' principle for  the operators under consideration.

In order to construct an appropriate scale of distributions  ({\it Riesz convolution algebra}) in Minkowski space $ \M $ 
we consider (for a fixed $ \xi \in \M $) a holomorphic ${\cal D}'$-valued mapping $\ {\bf C} \to {\cal D}'(\M)\ , \
\lambda \mapsto R_{\lambda}(x,\xi) $, such that $  R_{\lambda}(x,\xi) $ is an analytic continuation (in $ \lambda $) of the 
following (regular) distribution:
\be
\la{24}
\langle R_{\lambda}(x,\xi), g(x) \rangle =
\int\limits_{J_{+}(\xi)}^{}
\frac{\gamma(x,\xi)^{\lambda - \frac{n+1}{2}}}
{H_{n+1}(\lambda)}\, g(x)\, dx\ ,
\quad \mbox{\rm Re}\,\lambda > \frac{n-1}{2}\ ,
\ee
where $ dx = dx^0 \wedge dx^1 \wedge \ldots \wedge dx^n $ is a volume form in $ \M $, 
$ g(x) \in {\cal D}(\M) $, and $ H_{n+1}(\lambda) $
is a constant given by 
\be
\la{25}
H_{n+1}(\lambda) = 2 \pi^{\frac{n-1}{2}} 4^{\lambda - 1} 
\Gamma(\lambda) \Gamma\left(\lambda -
(n-1)/2\right)\ .
\ee

The following properties of this family of distributions are deduced directly from their definition.

For all $ \lambda \in {\bf C} $ and $ \xi \in \M $ we have
\be
\la{26}
\mbox{supp}\ R_{\lambda}(x, \xi) \subseteq \overline{J_{+}(\xi)} 
\ee
\be
\la{27}
\Box_{n+1}\, R_{\lambda} = R_{\lambda - 1}\ ,
\ee
\be
\la{28}
R_{\lambda} * R_{\mu} = R_{\lambda + \mu}\ ,\quad \mu \in {\bf C}\ , 
\ee
\be
\la{29}
(x - \xi, \partial_{x}) R_{\lambda} =  (2 \lambda - n + 1) R_{\lambda}\ ,
\ee
\be
\la{30}
\gamma^{\nu} R_{\lambda} =  4^{\nu} \left(\lambda \right)_{\nu} \left(\lambda - (n-1)/2\right)_{\nu} 
R_{\lambda + \nu}\ ,\quad  \nu \in {\bf Z}_{\geq 0}\ ,
\ee
where $ (\kappa)_{\nu} := \Gamma(\kappa + \nu) / \Gamma(\kappa) $ is Pochhammer's symbol, and 
$ \gamma = \gamma(x,\xi) $ is a square of the geodesic distance between $ x $ and $ \xi $ in $ \M $.

In addition, when $ n $ is odd, one can prove that
\be
\la{31}
R_{\lambda}(x, \xi) = \frac{1}{2 \pi^{\frac{n-1}{2}}}
\frac{ \delta_{+}^{ (\frac{n-1}{2} -\lambda) } (\gamma)}
{ 4^{\lambda-1} (\lambda - 1)!} \quad  \mbox{\rm for}\quad   \lambda = 1,2, \ldots, (n-1)/2 \ ,
\ee
where $ \delta_{+}^{(m)} (\gamma) $ stands for the $m$-th derivative of Dirac's delta-measure
concentrated on the surface of the future-directed characteristic half-cone  $ \overline{C_{+}(\xi)} $.

Another important property of Riesz distributions is that
\be
\la{32}
R_{0}(x, \xi) = \delta(x - \xi)\ .
\ee

Formulas \eq{26}, \eq{27}, \eq{32} show that $ R_{\lambda}(x,\xi) $ is a Riesz kernel for the ordinary
wave operator $ \Box_{n+1} $. The property \eq{31} means precisely that in even-dimensional Minkowski
spaces $ \M $  ($ n $ is odd) Huygens'     principle holds  for sufficiently low powers of the wave operator
$ \Box^{d} \,,\, d \leq (n-1)/2 $.

Now we are able to construct the Hadamard-Riesz expansion for the Riesz kernel
of a general self-adjoint wave-type operator \eq{18} on $ \M $.

First, we have to find a sequence of two-point smooth functions $\ U_{\nu} := U_{\nu}(x, \xi) \in \
{\cal C}^{\infty}(\Omega \times \Omega)\ , \ \nu = 0,1,2 \ldots $, as a solution of the following
{\it transport equations}:
\be
\la{33} 
\left(x-\xi, \partial_{x} \right) U_{\nu}(x,\xi) + \nu U_{\nu}(x, \xi) = - \frac{1}{4}\, {\cal L} \left[U_{\nu-1}(x,\xi)\right]\ , 
\quad \nu \geq 1\ .
\ee
It is well-known (essentially due to \cite{Had}) that the differential-recurrence  system \eq{33} has a {\it unique} 
solution provided each $ U_{\nu} $ is required to be  bounded 
in the vicinity of the vertex of the characteristic cone and $ U_{0}(x,\xi) $  is fixed for a normalization, i.e.
$$
U_{0}(x,\xi) \equiv 1 \ , \qquad U_{\nu}(\xi,\xi) \sim {\cal O}(1)\ ,\quad \forall\, \nu = 1,2,3, \ldots  
$$
These functions $ U_{\nu} $ are called {\it Hadamard's coefficients} of the operator
$ {\cal L}.$ 

In terms of $ U_{\nu} $  the required asymptotic expansion can be presented as follows:
\be
\la{34}
\Phi_{\lambda}^{\Omega}(x,\xi) \sim \sum\limits_{\nu=0}^{\infty} 
4^{\nu} (\lambda)_{\nu}\, U_{\nu}(x,\xi)\, R_{\lambda + \nu}(x,\xi) \ .
\ee
One can prove that for a hyperbolic differential  operator $ {\cal L} $ with locally analytic coefficients the Hadamard-Riesz
expansion is locally uniformly convergent. From now on we will restrict our consideration to this case.

For $ \lambda =1 $ formula \eq{34} provides an expansion of the fundamental solution of the operator  $ {\cal L} $ in  a
neighborhood of the vertex $ x =\xi $ of the characteristic cone:
\be
\la{35}
\Phi_{+}(x,\xi) = \sum\limits_{\nu=0}^{\infty} 
4^{\nu} {\nu}!\, U_{\nu}(x,\xi)\, R_{\nu +1}(x,\xi) \ .
\ee
When $ n $ is even, we have $\ \mbox{supp}\ R_{\nu+1}(x, \xi) = \overline{J_{+}(\xi)}\ $ for all $\, \nu = 0,1,2, \ldots $, and
therefore Huygens'     principle never occurs in odd-dimensional Minkowski spaces $ {\bf M}^{2l+1} $.

On the other hand, in the case of an odd number of space dimensions $ n \geq 3 $, we know due to \eq{31} that for
$ \nu = 0, 1, 2, \ldots, (n-3)/2 , \  \mbox{supp}\ R_{\nu+1}(x, \xi) = \overline{C_{+}(\xi)}\ $. Hence, using \eq{30}, we
can rewrite the series \eq{35} in following form:
\be
\la{36}
\Phi_{+}(x,\xi) = \frac{1}{2\pi^{p}} \left(  V(x,\xi)\, \delta_{+}^{(p-1)}(\gamma) + W(x,\xi)\, \eta_{+}(\gamma) \right)\ ,
\ee
where $ p := (n-1)/2 \ ,\ \eta_{+}(\gamma) $ is a regular distribution characteristic for the region $ J_{+}(\xi) $:
$$
\langle \eta_{+}(\gamma), g(x) \rangle = \int\limits_{J_{+}(\xi)}^{} g(x)\, dx\ , \quad g(x) \in {\cal D}(\M)\ ,
$$
and $ V(x,\xi) $ , $ W(x,\xi) $ are  analytic functions in a neighborhood of the vertex $ x=\xi $ which
admit the following expansions therein:
\be
\la{37}
V(x,\xi) =  \sum\limits_{\nu=0}^{p-1} \frac{1}{(1-p) \ldots (\nu-p)} \, U_{\nu}(x,\xi)\, \gamma^{\nu}\ ,
\ee
\be
\la{38}
W(x,\xi) =  \sum\limits_{\nu=p}^{\infty} \frac{1}{ (\nu-p)! }\, U_{\nu}(x,\xi)\, \gamma^{\nu - p} \ , \quad  p = \frac{n-1}{2}\ . 
\ee
The function $ W(x,\xi) $ is usually called {\it a logarithmic term} of the fundamental solution\footnote{
Such a terminology goes back to Hadamard's book \cite{Had}, where the function  $ W(x,\xi) $  is introduced
as a coefficient under the logarithmic singularity of an elementary solution (see for details \cite{Cou}, pp. 740--743).}.

It follows directly from the representation formula \eq{36} that operator $ {\cal L } $ satisfies Huygens'     principle
in a neighborhood of the point $ \xi $, if and only if, the logarithmic term $ W(x,\xi) $ of its fundamental solution
vanishes in this neighborhood identically in $ x $:\ $ W(x,\xi) \equiv 0\, $. 

The function $ W(x,\xi) $ is known to be a regular solution of the characteristic Goursat problem for the operator $ {\cal L}\, $:
\be
\la{39}
 {\cal L}\, \left[ W(x,\xi) \right]  = 0
\ee
with a boundary value given on the cone surface $ \overline{C_{+}(\xi)} \,$.  Such a boundary problem has a 
unique solution, and
hence, the necessary and sufficient condition for $ {\cal L } $ to be Huygens' operator becomes
\be
\la{40}
W(x,\xi)  \triangleq  0\ ,
\ee
where the symbol $ \triangleq $  implies that the equation in hand is satisfied only on $ \overline{C_{+}(\xi)} $\,.
By definition \eq{38}, the latter condition is equivalent to the following one
\be
\la{41}
U_{p}(x,\xi)  \triangleq  0\ ,\quad p = \frac{n-1}{2}\ .
\ee

In this way, we arrive at the important criterion for the validity of Huygens'     principle in terms of
coefficients of the Hadamard-Riesz expansion \eq{34}. Equation \eq{41} is essentially due to  
 Hadamard \cite{Had}. It will play a central role in the proof of our main theorem.

\section*{III. Proof of the main theorem}

We start with some  remarks concerning the properties of the one-dimensional Schr\"odinger operator
\be
\la{42}
L_{(k)} :=  -\left( \frac{\partial}{\partial \varphi}\right)^2 + v_{k}(\varphi)
\ee
with a general periodic soliton potential
\be
\la{43}
v_{k}(\varphi) := - 2 \left( \frac{\partial}{ \partial \varphi}\right)^2 \log 
{\cal W}\left[\Psi_{1}, \Psi_{2}, \ldots , \Psi_{N} \right] \ .
\ee
Here, as already discussed in the Introduction, $ {\cal W}\left[\Psi_{1}, \Psi_{2}, \ldots , \Psi_{N} \right] $ stands
for a Wronskian of the set of periodic functions on $ {\bf R}^{1} $:
\be
\la{44}
\Psi_{i}(\varphi) :=  \cos (k_{i}\, \varphi + \varphi_{i})\ , \quad  \varphi_{i} \in {\bf R}\ ,
\ee
associated to an arbitrary strictly monotonic sequence of real positive numbers ("soliton amplitudes"):
$\ 0 \leq k_1< \ldots < k_{N-1}<k_{N}\, $.

It is well-known (see, e.g., \cite{MS}) that any such operator $ L_{(k)} $ (as well as its proper 
solitonic counterpart \eq{16}) can be constructed by a successive application
of {\it Darboux-Crum  factorization transformations} (\cite{Darb}, \cite{Crum}) to the Schr\"odinger operator with 
the identically zero potential:
\be
\la{45}
L_{0} :=  -\left( \frac{\partial}{\partial \varphi}\right)^2 \ .
\ee

To be precise, let $ L $ be a second order ordinary differential operator with a sufficiently smooth potential:
\be
\la{46}
L :=  -\left( \frac{\partial}{\partial \varphi}\right)^2 + v(\varphi) \ .
\ee

We ask for formal factorizations of  the operator
\be
\la{47}
L - \lambda \, I = A^{*}\circ A\ ,  
\ee
where $ I $ is an identity operator, $\, \lambda \,$ is a (real) constant, and $\, A\, , \, A^{*}\, $ are the first order
operators adjoint to each other in a formal sense.

According to Frobenius' theorem (see, e.g., \cite{Ince}), the most general factorization \eq{47} is obtained  if we take
$ \chi(\varphi) $ as a generic element in $\, \mbox{Ker}(L - \lambda \, I )\setminus\left\{ 0 \right\}\, $
and set
\be
\la{48}
A := \chi \circ \left( \frac{\partial}{\partial \varphi} \right) \circ \chi^{-1}\quad , \quad 
A^{*} := - \chi^{-1} \circ \left( \frac{\partial}{\partial \varphi} \right) \circ \chi \ .   
\ee

Indeed, $\, A^{*}\circ A\, $ is obviously self-adjoint second order operator with the principal part
$\, - \partial^2 / \partial \varphi^2\, $.  Hence,  it is of the form \eq{46}. Moreover, since $\,  A[\chi] = 0 \,$,
we have $\,  \chi \in \mbox{Ker}\, A^{*}\circ A \,$, so that \eq{47} becomes  evident.

Note that for every $\, \lambda \in {\bf R}\, $  we actually get a one-parameter family of factorizations
of $\, L - \lambda \, I \,$. This follows from the fact that $\,  \dim\, \mbox{Ker}(L - \lambda \, I ) = 2 $,
whereas $ \chi(\varphi) $  and  $\, C\,\chi(\varphi) $ give rise to the same factorization pair $ (A, A^{*})\, $.

By definition, the Darboux-Crum transformation maps an operator $\, L = \lambda \, I + A^{*}\circ A \,$
into the operator 
\be
\la{49}
\tilde{L} :=  \lambda \, I + A \circ A^{*}\ ,  
\ee
in which $ A $ and $ A^{*} $ are interchanged. The operator $ \tilde{L} $ is also a (formally) self-adjoint 
second-order differential operator
\be
\la{50}
\tilde{L} :=  -\left( \frac{\partial}{\partial \varphi}\right)^2 + \tilde{v}(\varphi) \ ,
\ee
where $ \tilde{v}(\varphi) $ is given explicitly by
\be
\la{51}
\tilde{v}(\varphi)  =   v(\varphi) -  2 \left( \frac{\partial}{\partial \varphi}\right)^2 \log\, \chi(\varphi)  \ .
\ee
The initial operator $ L $ and its Darboux-Crum transform $ \tilde{L} $ are  obviously related to each other
via the following intertwining indentities:
\be
\la{52}
\tilde{L} \circ A = A \circ L \quad , \quad  L \circ A^{*} = A^{*} \circ \tilde{L} \ .
\ee

The Darboux-Crum transformation has a lot of important applications in the spectral theory of
Sturm-Liouville operators and related problems of quantum mechanics \cite{IH}.
In particular, it is used to insert or remove one eigenvalue without changing the rest of  the
spectrum of a Schr\"odinger operator (for details see the monograph \cite{MS} and 
 references therein).

The explicit construction  of the family of operators \eq{42} with periodic soliton potentials \eq{43}
is based on the following Crum's lemma:

\vspace{0.3cm}

{\bf Lemma}\,(\cite{Crum}).\ Let $ L $ be a given second order Sturm-Liouville operator \eq{46}
with a sufficiently smooth potential, and let $\, \left\{  \Psi_{1}, \Psi_{2}, \ldots  , \Psi_{N} \right\}\,$
be its eigenfunctions corresponding to arbitrarily fixed pairwise different eigenvalues 
$\, \left\{  \lambda_{1}, \lambda_{2},  \ldots  , \lambda_{N} \right\}\,$, i.e. $ \,   
\Psi_{i} \in \mbox{Ker} (L - \lambda_{i}\,I ) \, , \, i =1,2, \ldots , N \, $. Then, for arbitrary  
$\, \Psi \in \mbox{Ker} (L - \lambda\,I ) \, ,\, \lambda \in {\bf R}\,$, the function
\be
\la{53}
\chi_{N} (\varphi) := \frac{{\cal W}\left[\Psi_{1}, \Psi_{2}, \ldots , \Psi_{N}, \Psi \right]}
{{\cal W}\left[\Psi_{1}, \Psi_{2}, \ldots , \Psi_{N} \right]}
\ee
satisfies the differential equation 
\be
\la{54}
\left[ -\left( \frac{\partial}{\partial \varphi}\right)^2 + v_{N}(\varphi)\right]\,\chi_{N} (\varphi) = \lambda\,\chi_{N} (\varphi)
\ee
with the potential 
\be
\la{55}
 v_{N}(\varphi) :=  v(\varphi) - 2 \left( \frac{\partial}{\partial \varphi}\right)^2 
\log{\cal W}\left[\Psi_{1}, \Psi_{2}, \ldots , \Psi_{N} \right] \ .
\ee
\vspace{0.3cm}

Given a sequence of real positive numbers $ \, (k_{i})^{N}_{i=1} $: $\, 0 \leq k_1 < k_2 < \ldots < k_{N}\,$,
the Darboux-Crum factorization scheme:
\be
\la{56}
L_{i} := A_{i-1} \circ  A^{*}_{i-1} + k_{i}^{2}\,I =  A^{*}_{i} \circ  A_{i} + k_{i+1}^{2}\,I \  \rightarrow \ 
L_{i+1} := A_{i} \circ  A^{*}_{i} + k_{i+1}^{2}\,I \ ,
\ee
starting from the Schr\"odinger operator \eq{45} with a zero potential
$$
L_{0} \equiv -\left( \frac{\partial}{\partial \varphi}\right)^2 = A^{*}_{0} \circ  A_{0} + k_{1}^{2}\,I\ , 
$$
produces the required operator $ L_{(k)} \equiv L_{N} $ with the general periodic potential \eq{43}.

Now we  proceed to the proof of our main theorem formulated in the Introduction.

When  $\, N=0\, $, the statement of the theorem is evident, since the operator $ {\cal L}_{0} $ is just 
the ordinary wave operator in an odd number $ n $ of spatial variables.
 
Using  the Darboux-Crum scheme as outlined above we will carry out the proof by induction in $ N $.

Suppose that the statement of the theorem is valid for all $ m= 0, 1, 2, \ldots, N $. Consider an arbitrary
{\it integer} monotonic partition  $ (k_i) $ of height $ N $~: $\, 0<k_1< k_2 <  \ldots < k_{N}\, , \  
k_{i} \in {\bf Z} $. 

By our assumption, the  wave-type operator 
\be
\la{57}
{\cal L}_{N} := {\cal L}_{(k)} = \Box_{n+1} + u_{k}(x)\ ,
\ee
associated to this partition, satisfies Huygens'     principle in the $(n+1)$-dimensional Minkowski space $ \M $
with $ n $ odd, and $\,  n \geq 2 \, k_{N} + 3 \,$.  We fix the minimal admissible  number of space variables,
i.e. $ \, n = 2 \, k_{N} + 3 \, $, and denote
\be
\la{58}
 p:= \frac{n-1}{2} = k_{N} + 1\ .
\ee

By construction, the operator $ {\cal L}_{N} $ can be written explicitly in terms of suitably chosen
cylindrical coordinates in $ \M $:
\be
\la{59}
{\cal L}_{N} = \Box_{n-1} - \left[ 
\left( \frac{\partial}{\partial r}\right)^{2} + \frac{1}{r} \, \frac{\partial}{\partial r} - \frac{1}{r^2}\,\left( 
 -\left( \frac{\partial}{\partial \varphi}\right)^2 + v_{N}(\varphi) \right)
\right]\ ,
\ee
where $ (r, \varphi) $ are the polar coordinates in some Euclidean $2$-plane $ E $ orthogonal to the
time direction in $ \M $, i.e. $\, E \in {\bf Gr}_{ \perp} (n+1,2)\,$; $\, \Box_{n-1}\, $  is a wave operator in the 
orthogonal  complement $ \, E^{\perp} \cong  {\bf M}^{n-1} \, $ of $ E $  in $ \M $; and $ v_{N}(\varphi) $ is 
a $ 2\pi$-periodic potential given by \eq{43}.

Let $ k := k_{N+1} $ be an arbitrary positive integer such that
\be
\la{60}
k > k_{N}\ .
\ee  

We apply the Darboux-Crum transformation \eq{56} with the spectral parameter $ k $ to the angular
part of the Laplacian in $ E $. For this we rewrite $ {\cal L}_{N} $ in the form 
\be
\la{61}
{\cal L}_{N} = \Box_{n-1} - \left[ 
\left( \frac{\partial}{\partial r}\right)^{2} + \frac{1}{r} \, \frac{\partial}{\partial r} - \frac{1}{r^2}\,\left( 
A_{N}^{*} \circ A_{N} + k^2  \right)
\right]\ ,
\ee
and set
\be
\la{62}
{\cal L}_{N+1} := \Box_{n-1} - \left[ 
\left( \frac{\partial}{\partial r}\right)^{2} + \frac{1}{r} \, \frac{\partial}{\partial r} - \frac{1}{r^2}\,\left( 
A_{N} \circ A^{*}_{N} + k^2  \right)
\right]\ ,
\ee
where $\,  A_{N} := A_{N}(\varphi) \,$ and $\, A^{*}_{N} := A^{*}_{N}(\varphi) \,$
are the first order ordinary differential operators of the  form \eq{48}.

According to \eq{52}, we have
\be
\la{63}
{\cal L}_{N+1} \circ A_{N} = A_{N} \circ {\cal L}_{N} \quad , \quad {\cal L}_{N} \circ A_{N}^{*} = A_{N}^{*} \circ {\cal L}_{N+1} \ .
\ee

Let $\, \Phi_{\lambda}^{N}(x,\xi)\, $ and $\, \Phi_{\lambda}^{N+1}(x,\xi)\, $  be the Riesz kernels of  hyperbolic operators
$ {\cal L}_{N}  $ and $ {\cal L}_{N+1} $ respectively. Then, by virtue of \eq{63} we must  have the relation
\be
\la{64}
A^{*}_{N}(\varphi) \, \left[  \Phi_{\lambda}^{N+1} \right] - A_{N}(\phi) \, \left[  \Phi_{\lambda}^{N} \right] = 0 
 \quad \mbox{for all}\ \lambda \in {\bf C}\ ,
\ee
where $ \,  A_{N}(\phi)  \,$ is the differential operator  $\, A_{N} \,$ written in terms of the variable $ \phi $ conjugated
to $\, \varphi\, $.  Indeed, if  identity \eq{64}  were not  valid, one could  define a holomorphic mapping
$ \tilde{\Phi}^{N}\,:\, {\bf C} \to {\cal D}',\, \lambda \mapsto \tilde{\Phi}_{\lambda}^{N}(x,\xi)\,$, such that
\be
\la{65}
\tilde{\Phi}_{\lambda}^{N}(x,\xi) := {\Phi}_{\lambda}^{N}(x,\xi) + a \, \left(
A^{*}_{N}(\varphi) \, \left[  \Phi_{\lambda}^{N+1} \right] - A_{N}(\phi) \, \left[  \Phi_{\lambda}^{N} \right] \right)\ .
\ee
The distribution $\,\tilde{\Phi}_{\lambda}^{N}(x,\xi) \,$,  depending on an arbitrary complex parameter $ a \in {\bf C} $,
would also satisfy all the axioms \eq{21} in the definition of a Riesz kernel for the operator  $ {\cal L}_{N} $.
In this way, we would arrive at the contradiction with the uniqueness of such a kernel.

In particular, when $ \lambda = 1 $, the identity \eq{64}  gives the relation between the fundamental solutions
$\,  {\Phi}_{+}^{N}(x,\xi) \equiv {\Phi}_{1}^{N}(x,\xi) \, $ and $\,  {\Phi}_{+}^{N+1}(x,\xi) \equiv {\Phi}_{1}^{N+1}(x,\xi) \, $ 
of operators  $ {\cal L}_{N} $ and  $ {\cal L}_{N+1} \,$.  In accordance with \eq{36}, we have
\be
\la{66}
\Phi^{N}_{+}(x,\xi) = \frac{1}{2\pi^{p}} \left(  V_{N}(x,\xi)\, \delta_{+}^{(p-1)}(\gamma) + W_{N}(x,\xi)\, \eta_{+}(\gamma) \right)
\ee
and
\be
\la{67}
\Phi^{N+1}_{+}(x,\xi) = \frac{1}{2\pi^{p}} \left(  V_{N+1}(x,\xi)\, \delta_{+}^{(p-1)}(\gamma) + W_{N+1}(x,\xi)\, \eta_{+}(\gamma) \right)\ ,
\ee
where $ \gamma $ is a square of the geodesic distance between the points $ x $ and $ \xi $ in $ \M $. Substituting \eq{66}, \eq{67}
into \eq{64}, we get the relation between the logarithmic terms $ \, W_{N}(x,\xi)\, $  and $\, W_{N+1}(x,\xi)\, $ of operators
$ {\cal L}_{N} $ and $ {\cal L}_{N+1}\, $
\be
\la{68}
A^{*}_{N}(\varphi) \, \left[  W_{N+1}(x,\xi) \right] - A_{N}(\phi) \, \left[  W_{N}(x,\xi) \right] = 0 \ .
\ee
By our assumption, $ {\cal L}_{N} $  is a Huygens' operator in $ \M $, so that $\, W_{N}(x,\xi) \equiv 0 $. Hence, equation
\eq{68} implies $\, A^{*}_{N}(\varphi) \, \left[  W_{N+1}(x,\xi) \right] = 0 \, $. On the other hand, as discussed in Sect.~II, 
the logarithmic term $\, W_{N+1}(x,\xi) \,$  is a regular solution of the characteristic Goursat problem for $ {\cal L}_{N+1}\, $,
i.e. in particular,
\be
\la{69}
{\cal L}_{N+1}\, \left[ W_{N+1}(x,\xi) \right] = 0 \ .
\ee
Taking into account definition \eq{62} of the operator  $ {\cal L}_{N+1}\, $, we arrive at the following equation for
$\, W_{N+1}(x,\xi) \,$:
\be
\la{70}
 \Box_{n-1} W_{N+1}(x,\xi) =  
\left( \left(\frac{\partial}{\partial r}\right)^{2} + \frac{1}{r} \, \frac{\partial}{\partial r} - \frac{k^2}{r^2} \right)\,W_{N+1}(x,\xi) \ .
\ee
According to \eq{38},  the logarithmic term $ W_{N+1} $ admits the following expansion 
\be
\la{71}
W_{N+1}(x,\xi) =  \sum\limits_{\nu=p}^{\infty} \, U_{\nu}(x,\xi)\, \frac{\gamma^{\nu - p} }{ (\nu-p)!}  \ , \quad  p = \frac{n-1}{2}\ , 
\ee
where $ \, U_{\nu}(x,\xi)\, $ are the Hadamard coefficients of the operator $ \, {\cal L}_{N+1}\,$. Since the potential of the wave-type
operator $ \, {\cal L}_{N+1}\,$ depends only on the variables $\, r, \varphi \,$, its Hadamard coefficients $\, U_{\nu}\, $
must depend on the same variables $\, r, \varphi \, $ and their conjugates $\, \rho, \phi \,$ only:
\be
\la{72}
U_{\nu} = U_{\nu}(r, \varphi, \rho, \phi) \qquad \mbox{for all}\quad \nu= 0, 1, 2, \ldots  
\ee
This follows immediately from the uniqueness of solution of Hadamard's transport equations \eq{33}.

On the other hand, since
\be
\la{73}
\gamma = s^2 - r^2 - \rho^2 + 2 r\,\rho\, \cos(\varphi - \phi)\ ,
\ee
where $ s $ is a geodesic distance in the space $ \, E^{\perp} \cong {\bf M}^{n-1} \, $ orthogonally
complementary to the  $2$-plane $ E $, we conclude that $  W_{N+1} $ is actually a function of five
variables: $ \, W_{N+1} =  W_{N+1}(s, r, \rho, \varphi, \phi )\, $.  On the space of such functions the wave 
operator  $\,  \Box_{n-1} \, $ in $ E^{\perp} $ acts in the same way as its ``radial part'', i.e.
$$
 \Box_{n-1} W_{N+1} =   
\left( \left(\frac{\partial}{\partial s}\right)^{2} + \frac{n-2}{s} \, \frac{\partial}{\partial s} \right)\,W_{N+1} \ . 
$$
Hence, equation \eq{70} becomes
\be
\la{74}
\left( \left(\frac{\partial}{\partial r}\right)^{2} - \left(\frac{\partial}{\partial s}\right)^{2}  - \frac{n-2}{s} \, \frac{\partial}{\partial s} 
+ \frac{1}{r} \, \frac{\partial}{\partial r} - \frac{k^2}{r^2} \right)\,W_{N+1} = 0 \ .
\ee
Now we substitute the expansion \eq{71}
\be
\la{75}
W_{N+1} =  \sum\limits_{\nu=p}^{\infty} \, U_{\nu}(r, \varphi, \rho, \phi)\, \frac{ \gamma^{\nu - p}}{(\nu-p)!} \ , 
\quad  p = \frac{n-1}{2}\ , 
\ee
into the left-hand side of the latter equation and develop the result into the similar power series in $ \gamma \,$, 
taking into account formula \eq{73}. After simple calculations we obtain
\be
\la{76}
\sum\limits_{\nu=p}^{\infty}\, 
\Biggl[ 
\left( U_{\nu}'' + \frac{1}{r}\, U_{\nu}' - \frac{k^2}{r^2} U_{\nu} \right) 
- 4 \,\left( r - \rho\,\cos(\varphi -\phi) \right)\, U_{\nu+1}' -  
\ee
$$
-\, 2\, \left(  2\,(\nu+1) - \frac{\rho}{r}\, \cos(\varphi -\phi)\right)\,U_{\nu+1} 
- 4\,\rho^{2}\,\sin^{2}(\varphi -\phi)\,U_{\nu+2}
\Biggr]\, 
\frac{ \gamma^{\nu-p}}{(\nu-p)!} = 0\ ,
$$
where the prime means differentiation with respect to $ r \,$. 

Since the  functions $ U_{\nu} $ do not depend  explicitly on $ \gamma \,$, equation \eq{76} can be satisfied
only if each coefficient under the powers of $ \gamma $ vanishes separately. In this way we arrive at the following
differential-recurrence relation for the Hadamard coefficients of the operator  $ {\cal L}_{N+1}\,$:
$$
4\,\rho^{2}\,\sin^{2}(\varphi -\phi)\,U_{\nu+2} = \left( U_{\nu}'' + \frac{1}{r}\, U_{\nu}' - \frac{k^2}{r^2} U_{\nu} \right)  + 
$$
\be
\la{77}
+\,  \frac{2\rho}{r}\, \cos(\varphi -\phi)\, \left( 2\,r\,  U_{\nu+1}' +  U_{\nu+1} \right)  
- 4\, \left( r\, U_{\nu+1}' + (\nu +1)\,U_{\nu+1} \right)\ ,
\ee
where $ \nu $ runs from $ p\, $: $\, \nu = p,\, p+1,\, p+2, \ldots $

 To get a further simplification of equation \eq{77} we notice that all the Ha\-da\-mard coefficients of the operators
under consideration \eq{11}, \eq{10} are homogeneous functions of appropriate degrees. More precisely, they
have the following  specific form 
\be
\la{78}
U_{\nu} (r, \varphi, \rho, \phi) = \frac{1}{(r\,\rho)^{\nu}}\, \sigma_{\nu} (\varphi, \phi) \ , \quad  \nu = 0, 1, 2, \ldots\ ,
\ee
where $\, \sigma_{\nu} (\varphi, \phi) = \sigma_{\nu} (\phi, \varphi) \, $  are symmetric $2\pi$-periodic functions
depending on the angular variables only.

In order to prove Ansatz \eq{78} we have to go back to the relation \eq{64} between  the Riesz kernels of operators
$ {\cal L}_{N} $ and $ {\cal L}_{N+1}\,$:
\be
\la{79}
A^{*}_{N}(\varphi) \, \left[  \Phi_{\lambda}^{N+1}(x,\xi) \right] - A_{N}(\phi) \, \left[  \Phi_{\lambda}^{N} (x,\xi) \right] = 0 \ ,
 \quad  \lambda \in {\bf C}\ ,
\ee
If we substitute the Hadamard-Riesz expansions \eq{34} of the kernels  $\, \Phi_{\lambda}^{N} (x,\xi)\, $
and  $\, \Phi_{\lambda}^{N+1} (x,\xi)\, $ into \eq{79} directly and take into account that $ A_{N} $ and its 
adjoint $ A^{*}_{N} $ are the first order ordinary differential operators of the following form (cf. \eq{48}):
\be
\la{80}
A_{N}(\varphi) = \frac{\partial}{\partial \varphi} - f_{N}(\varphi)\ , \quad 
A_{N}^{*}(\varphi) = - \frac{\partial}{\partial \varphi} - f_{N}(\varphi)\ , 
\ee
where $\, f_{N}(\varphi) = (\partial/\partial \varphi) \log \chi_{N}(\varphi)\,$, we obtain
$$
\sum\limits_{\nu = 0}^{\infty}\, 4^{\nu}\, (\lambda)_{\nu}\, \biggl[ 2 r \rho\,\sin(\varphi - \phi)\, 
\left( U^{N+1}_{\nu+1} -  U^{N}_{\nu+1} \right) - 
$$
\be
\la{81}
-\, \left(  \frac{\partial}{\partial \varphi} + f_{N}(\varphi)\right)\, U^{N+1}_{\nu} -
\left(  \frac{\partial}{\partial \phi} - f_{N}(\phi)\right)\, U^{N}_{\nu} \biggr]\, R_{\lambda + \nu } = 0 \ ,
\ee
where $\, U_{\nu}^{N}(r, \varphi, \rho, \phi)\,$ and $\, U_{\nu}^{N+1}(r, \varphi, \rho, \phi)\,$ are the Hadamard
coefficients of operators $ {\cal L}_{N} $ and $ {\cal L}_{N+1} $ respectively;  $\, R_{\lambda} := 
R_{\lambda}(x,\xi) $ is the family of Riesz  distributions in  $ \M \,$.

The same argument as above (see the remark before formula \eq{77}) shows that all the coefficients of
 the series \eq{81} under the Riesz distributions of different weights must vanish separately. So we arrive
at the recurrence relation  between the sequences of Hadamard's coefficients of operators $ {\cal L}_{N} $
 and  $ {\cal L}_{N+1}\,$:
\be
\la{82}
U^{N+1}_{\nu+1}  = U^{N}_{\nu+1}  + \frac{1}{2 r \rho\,\sin(\varphi - \phi)}\,
\left[ 
\left(  \frac{\partial}{\partial \varphi} + f_{N}(\varphi)\right)\, U^{N+1}_{\nu}  +
\left(  \frac{\partial}{\partial \phi} - f_{N}(\phi)\right)\, U^{N}_{\nu}
\right]\ ,
\ee
where $ \, U^{N+1}_{0} = U^{N}_{0}  \equiv 1\, $ and $ \nu = 0, 1, 2, \dots $ 
Now it is easy to conclude  from \eq{82} by induction in $ N $ that the Ansatz
\eq{78} really holds  for Hadamard's coefficients of all wave-type operators \eq{11} 
with potentials \eq{10}.

Returning to equation \eq{77} and substituting \eq{78} therein, we obtain the following
three-term recurrence relation for the angular functions $ \sigma_{\nu}(\varphi, \phi)\,$:
\be
\la{83}
4\,\sin^{2}(\varphi - \phi)\, \sigma_{\nu+2} = (\nu^2 - k^2)\,\sigma_{\nu} - 
2 (2\nu+1)\, \cos(\varphi - \phi)\,\sigma_{\nu+1} \ , 
\ee
where $\, \nu = p,\,p+1,\, p+2, \ldots \,$.

In order to analyze equation \eq{83} it is convenient to introduce a  formal generating function
for the quantities $ \{\, \sigma_{\nu}\,\}\,$:
\be
\la{84}
F(t) := \sum\limits_{\nu = p}^{\infty} \,\sigma_{\nu}(\varphi, \phi)\, \frac{t^{\nu-p}}{(\nu - p)!} \ .
\ee

The recurrence relation \eq{83} turns out to be  equivalent to the classical hypergeometric
differential equation for the function $ F(t) $
\be
\la{85}
\left( 4\,(1-\omega^{2}) + 4 \omega t - t^2 \right)\, \frac{d^2 F}{d t^2} + (2 p + 1)\,(2 \omega - t )\,\
\frac{d F}{d t} + (k^2 - p^2)\,F = 0 \ ,
\ee
where $ \omega := \cos(\varphi - \phi)\, $.  The general solution to \eq{85} is given in terms
of Gauss' hypergeometric series:
\be
\la{86}
F(t) = C\, {}_{2}\,{\bf F}_{1} (p-k; p+k; p+1/2\, |\, z) + C_{1}\, z^{-p+1/2}\,{}_{2}\,{\bf F}_{1} (1/2-k; 1/2+k; 3/2 - p \, | \, z) \ ,
\ee
where $ z :=  (t- 2 \omega +2 )/4 $ and  $\, {}_{2}\,{\bf F}_{1}\,$ is defined by 
\be
\la{87}
{}_{2}\,{\bf F}_{1} (a; b; c\, |\, z ) := \sum\limits_{\mu=0}^{\infty}\, \frac{(a)_{\mu} (b)_{\mu}}{(c)_{\mu}}\, \frac{z^{\mu}}{\mu!}\ .
\ee

As discussed in Sect.II, the Hadamard coefficients $ U_{\nu}(x,\xi) $  must be regular in a neighborhood of the vertex
of the characteristic cone $ x = \xi\, $. When $\, x \to \xi \,$, we have $\, \omega \to 1\, $ and $\, U_{p}(\xi,\xi) \propto 
\sigma_{p}(\phi, \phi) = F(0) |_{\omega =1}\, $  is not bounded unless $\, C_{1} = 0\, $. 

In this way,  setting $\, C_{1} = 0\, $ in \eq{86},  we obtain	
\be
\la{88}
\sum\limits_{\nu = p}^{\infty} \,\sigma_{\nu}(\varphi, \phi)\, \frac{t^{\nu-p}}{(\nu - p)!} =  
 C\  {}_{2}\,{\bf F}_{1} \left( p-k; p+k; p+1/2\, |\,  (t- 2 \omega + 2 )/4 \right) \ .
\ee
Now it remains to recall that by our assumption \eq{60} $\, k \in {\bf Z}\,$ and $\, k > k_{N} \,$. Since $\, 
p = (n-1)/2 = k_{N} +1\,$,  we have $\, k \geq p\, $. So  the  hypergeometric  series in the right-hand side of equation \eq{88}
is truncated. In fact, the generating function  \eq{84} is expressed in terms of the classical Jacobi polynomial
$ \, {\bf P}^{(p-1/2 ,  p+1/2)}_{k-p} (\omega - t/2)\, $ of degree $ \, k-p \,$. Hence, $ \,  \sigma_{k+1} (\varphi, \phi) \equiv  0 \,$,
and the $(k+1)$-th Hadamard coefficient of the operator $ {\cal L}_{N+1} $ vanishes identically:
\be
\la{89}
U_{k+1}(x, \xi) \equiv 0 \ .
\ee 
According to Hadamard's criterion \eq{41},  it means that the operator $ {\cal L}_{N+1} $ satisfies Huygens'  
    principle in Minkowski space $ \M \,$, if $ n $ is odd and 
$$
n \geq 2\,k + 3\ ,
$$
Thus, the proof of the theorem is completed.

\section*{IV. Concluding remarks and examples}

\par In the present paper we have constructed a new hierarchy of Huygens' operators
in higher dimensional Minkowski spaces $ \M , n > 3 $. However, the problem of
complete description of the whole class of such operators for arbitrary $ n $ still
remains open. As mentioned in the Introduction, the famous Hadamard's conjecture
claiming that any Huygens' operator $ {\cal L} $ can be reduced to the ordinary
d'Alembertian $ \, \Box_{n+1} \,$ with the help of trivial transformations is valid
only in $ {\bf M}^{3+1}\,$. Recently, in the work \cite{Ber} one of the authors put 
forward the relevant modification of Hadamard's  conjecture for Minkowski spaces
of arbitrary dimensions. Here we recall and discuss briefly this statement.

Let $ \Omega $ be an open set in Minkowski space $ \M \cong {\bf R}^{n+1}\, $,
and let $\, {\cal F}(\Omega)\, $ be a ring of partial differential operators defined over the
function space $\, C^{\infty}(\Omega)\,$. For a fixed pair of operators $\, {\cal L}_{0}, 
{\cal L} \in {\cal F}(\Omega)\, $ we introduce the map 
\be
\la{90}
 \mbox{\rm ad}_{{\cal L}, {\cal L}_{0}}:\ {\cal F}(\Omega) \to 
{\cal F}(\Omega) \ , \quad  A \mapsto \mbox{\rm ad}_{{\cal L}, {\cal L}_{0}}[A] \ ,
\ee
such that
\be
\la{91}
 \mbox{\rm ad}_{{\cal L}, {\cal L}_{0}}[A] := {\cal L} \circ A - A \circ {\cal L}_{0}\ . 
\ee
Then, given $\, M \in {\bf Z}_{>0}\, $, the iterated $ \mbox{\rm ad}_{{\cal L}, {\cal L}_{0}}$-map 
is determined by
\be
\la{92}
\mbox{\rm ad}_{{\cal L}, {\cal L}_{0}}^{M}[A] := 
\mbox{\rm ad}_{{\cal L}, {\cal L}_{0}}\left[
\mbox{\rm ad}_{{\cal L}, {\cal L}_{0}}\left[
\ldots \mbox{\rm ad}_{{\cal L}, {\cal L}_{0}}[A]\right]\ldots
\right] = 
\sum\limits_{k=0}^{M} (-1)^{k} {M \choose k} {\cal L}^{M-k}\circ A \circ
{\cal L}_{0}^{k}\ .
\ee

\vspace{0.3cm}

{\bf Definition.}\ The operator $ \, {\cal L} \in  {\cal F}(\Omega)\, $ is called {\it M-gauge related}
to the operator $ \, {\cal L}_{0} \in  {\cal F}(\Omega)\, $, if there exists  a smooth function
$\, \theta(x) \in {\cal C}^{\infty}(\Omega)\,$ non-vanishing in $ \Omega $, and an integer positive number
$ M \in {\bf Z }_{>0}\,$,  such that
\be
\la{93}
\ad \left[ \theta(x) \right] \equiv 0 \quad \mbox{identically in}\ {\cal F}(\Omega)\ . 
\ee

In particular, when $ M = 1\, $, the operators   $ {\cal L} $ and $ {\cal L}_{0} $ are connected just by 
the trivial gauge transformation $\, {\cal L} = \theta(x) \circ {\cal L}_{0} \circ \theta(x)^{-1}\,$.

The modified Hadamard's conjecture  claims:

\vspace{0.2cm}

\begin{quote}
{\it Any Huygens' operator $ {\cal L} $ of the general form
\be
\la{94}
{\cal L} = \Box_{n+1} + \left( a(x), \partial \right) + u(x)\ ,
\ee
 in a Minkowski space $ \M $ ($ n $ is odd, $ n \geq 3 $) is $M$-gauge 
related to the ordinary wave operator $ \Box_{n+1} $ in $ \M $.
}

\end{quote}

\vspace{0.2cm}

For Huygens' operators associated to the rational solutions of the KdV-equation \eq{2}, \eq{3} and
to Coxeter groups \eq{5}, \eq{6} this conjecture has been proved in \cite{Ber} and \cite{BM}. In these 
cases the required identities \eq{93} are the following
\be
\la{95}
\mbox{\rm ad}_{{\cal L}_{k}, {{\cal L}_{0}}}^{M_{k}+1}[{\cal P}_{k}(x^0)] = 
0\ , \quad  M_{k} := \frac{k (k+1) }{2}\ ,
\ee
where $ {\cal L}_{k} $ is given by \eq{2} with the potential \eq{3} for $\, k=0,1,2, \ldots $ and
\be
\la{96}
\mbox{\rm ad}_{{\cal L}_{m}, {{\cal L}_{0}}}^{M_{m}+1}[{\pi}_{m}(x)] = 0\ , \quad 
 M_{m} := \sum\limits_{\alpha \in \Re_{+}}^{} m_{\alpha}\ ,
\ee
where $ {\cal L}_{m} $ is defined by \eq{5}, \eq{6} and $\, \pi_{m}(x) := 
\prod_{\alpha \in \Re_{+}}^{} (\alpha, x)^{m_{\alpha}}\, $.

It is remarkable that for the operators constructed in the present work
the modified Hadamard's conjecture is also verified. More precisely,
for a given wave-type operator
\be
\la{97}
{\cal L}_{(k)} = \Box_{n+1} - \frac{2}{r^2} \left( \frac{\partial}{ \partial \varphi}\right)^2 \log 
{\cal W}\left[\Psi_{1}(\varphi), \Psi_{2}(\varphi), \ldots , \Psi_{N}(\varphi) \right] \ ,
\ee
associated to a positive integer partition $ (k_{i})\,$:$\, 0 \leq k_1 < k_2 < \ldots < k_N\,$,
we have the identity
\be
\la{98}
\mbox{\rm ad}_{{\cal L}_{(k)}, {{\cal L}_{0}}}^{|k|+1}[ \Theta_{(k)}(x)] =  0\ , 
\ee
where $\,  \Theta_{(k)}(x) := r^{|k|}\, {\cal W}\left[\Psi_{1}, \Psi_{2}, \ldots , \Psi_{N} \right]  \,$
and $\,  |k| := \sum\nolimits_{i=1}^{N} k_{i} \,$ is a weight of the partition $\, (k_{i}) \,$.

We are not going to prove \eq{98} in the present paper.  A more detailed discussion 
of this identity and associated algebraic structures will be the subject of our subsequent
work. Here, we only mention that such type identities naturally appear \cite{Ber}--\cite{Ber1} 
in connection with a classification of overcomplete  commutative rings of partial 
differential operators
\cite{VCh1}, \cite{VCh2}, \cite{VChS}, and  with the bispectral problem \cite{DG}.

We conclude the paper with several concrete examples illustrating  our main theorem.

{\bf 1.}\  As a first example we consider the dihedral group $ I_{2}(q) ,\, q \in {\bf Z}_{>0}, $ acting on the 
Euclidean plane $ E \cong {\bf R^2} \subset {\bf Gr}_{\perp} (n+1,2) $ 
and fix the simplest partition $ k=(q) $ and the  phase $ \varphi = \pi/2 $.  According to Remark~II, 
in this case our theorem gives the wave-type operator with the Calogero-Moser potential related
to the Coxeter group $ I_{2}(q) $ with $ m=1 $:
$$
{\cal L}_{(k)} = \Box_{n+1} + \frac{2\,q^2}{r^2 \sin^{2}(q\,\varphi)}\ .
$$
This operator satisfies Huygens'     principle in $ \M $ if $ n $ is odd and $  n \geq 2\,q +3 $.
The Hadamard coefficients of   $ {\cal L}_{(k)} $ can be presented in a simple closed
form in terms of polar coordinates on $ E $:
$$
U_{0} =1\ ,
$$
$$
U_{\nu} = \frac{1}{(2r\rho)^{\nu}}\, \frac{T_{q}^{(\nu)}(\cos(\varphi - \phi))}{\sin(q\, \varphi) \sin(q\, \phi)}\ , \quad \nu \geq 1\ ,
$$
where $ T_{q}(z) := \cos(q\, \arccos(z)), \, z \in [-1,1] $,  is the $q$-th Chebyshev polynomial, and
$ T_{q}^{(\nu)}(z) $ is its derivative of order $ \nu $ with respect to $ z $. 
These formulas are easily obtained with the help of recurrence relation \eq{82}.

{\bf 2.}\  Now we fix $ N = 2 ,\, k_{1}=2,\, k_{2}=3 $ and $\, \varphi_{1}= \pi/2 ,\, \varphi_{2}=0\, $.
The corresponding wave-type operator  
$$
{\cal L}_{(k)} = \Box_{n+1} + 
\frac{ 
10\,\left( x_{1}^{2}+ x_{2}^{2} \right) 
\left(15\, x_{2}^{2}- x_{1}^{2}\right)
}
{
\left(5 x_{2}^{2}+ x_{1}^{2} \right)^{2}\, x_{1}^{2}
}\ ,
$$
satisfies Huygens'     principle for odd $ n \geq 9 $.  The nonzero Hadamard coefficients
of this operator are given explicitly by the formulas:
$$
U_{{0}}=1,
$$
$$
U_{{1}}={\frac {40\,x_{{
2}}\xi_{{1}}\xi_{{2}}x_{{1}}+15\,{\xi_{{1}}}^{2}{x_{{2}
}}^{2}+75\,{\xi_{{2}}}^{2}{x_{{2}}}^{2}+15\,{\xi_{{2}}}
^{2}{x_{{1}}}^{2}-5\,{\xi_{{1}}}^{2}{x_{{1}}}^{2}}{2\,
\xi_{{1}}x_{{1}}\left (5\,{x_{{2}}}^{2}+{x_{{1}}}^{2}
\right )\left (5\,{\xi_{{2}}}^{2}+{\xi_{{1}}}^{2}
\right )}},
$$
$$
U_{{2}}={\frac {120\,x_{{2}}\xi_{{1}}\xi_{{2
}}x_{{1}}+15\,{\xi_{{2}}}^{2}{x_{{1}}}^{2}-5\,{\xi_{{1}
}}^{2}{x_{{1}}}^{2}+15\,{\xi_{{1}}}^{2}{x_{{2}}}^{2}+75
\,{\xi_{{2}}}^{2}{x_{{2}}}^{2}}{4\,{\xi_{{1}}}^{2}{x_{{
1}}}^{2}\left (5\,{x_{{2}}}^{2}+{x_{{1}}}^{2}\right )
\left (5\,{\xi_{{2}}}^{2}+{\xi_{{1}}}^{2}\right )}},
$$
$$
U_{{3}}=-{\frac {15\,x_{{2}}\xi_{{2}}}{{\xi_{{1}}}^{2}{x_{
{1}}}^{2}\left (5\,{x_{{2}}}^{2}+{x_{{1}}}^{2}\right )
\left (5\,{\xi_{{2}}}^{2}+{\xi_{{1}}}^{2}\right )}}\ .
$$

{\bf 3.} Now we take $ N=3 $, the partition $ k = (1,\, 3,\, 4)\,$, 
and the phases $\, \varphi_{1}= \varphi_{2}= \varphi_{3}= \pi/2\, $. The corresponding
operator  
$$
{\cal L}_{(k)} =  \Box_{n+1} + 
\frac{ 
12\, \left(
49 x_{1}^{4}+ 28 x_{1}^{2} x_{2}^{2} - x_{2}^{4} \right)
}
{ 
x_{2}^{2}\,
\left (7 x_{1}^{2} + x_{2}^{2} 
\right)^{2}
},
$$
is a Huygens operator in $ \M $ when $ n $ is odd and $ n \geq 11\, $.
The nonzero Hadamard's coefficients are
$$
U_{{0}}=1,
$$
$$
U_{{1}}={\frac {-21\,{\xi_{{2}}}^{2}{x_{{1}}}^{2}-42\,x_
{{2}}\xi_{{1}}\xi_{{2}}x_{{1}}-21\,{\xi_{{1}}}^{2}{x_{{
2}}}^{2}+3\,{\xi_{{2}}}^{2}{x_{{2}}}^{2}-147\,{\xi_{{1}
}}^{2}{x_{{1}}}^{2}}{\xi_{{2}}x_{{2}}\left (7\,{x_{{1}}
}^{2}+{x_{{2}}}^{2}\right )\left (7\,{\xi_{{1}}}^{2}+{
\xi_{{2}}}^{2}\right )}},
$$
$$
U_{{2}}={\frac {735\,{\xi_{{1}
}}^{2}{x_{{1}}}^{2}+504\,x_{{2}}\xi_{{1}}\xi_{{2}}x_{{1
}}+105\,{\xi_{{1}}}^{2}{x_{{2}}}^{2}-21\,{\xi_{{2}}}^{2
}{x_{{2}}}^{2}+105\,{\xi_{{2}}}^{2}{x_{{1}}}^{2}}{4\,{
\xi_{{2}}}^{2}{x_{{2}}}^{2}\left (7\,{x_{{1}}}^{2}+{x_{
{2}}}^{2}\right )\left (7\,{\xi_{{1}}}^{2}+{\xi_{{2}}}^
{2}\right )}},
$$
$$
U_{{3}}={\frac {-1260\,x_{{2}}\xi_{{1}}
\xi_{{2}}x_{{1}}+21\,{\xi_{{2}}}^{2}{x_{{2}}}^{2}-105\,
{\xi_{{1}}}^{2}{x_{{2}}}^{2}-105\,{\xi_{{2}}}^{2}{x_{{1
}}}^{2}-735\,{\xi_{{1}}}^{2}{x_{{1}}}^{2}}{8\,{\xi_{{2}
}}^{3}{x_{{2}}}^{3}\left (7\,{x_{{1}}}^{2}+{x_{{2}}}^{2
}\right )\left (7\,{\xi_{{1}}}^{2}+{\xi_{{2}}}^{2}
\right )}},
$$
$$
U_{{4}}={\frac {315\,x_{{1}}\xi_{{1}}}{4\,{
\xi_{{2}}}^{3}{x_{{2}}}^{3}\left (7\,{x_{{1}}}^{2}+{x_{
{2}}}^{2}\right )\left (7\,{\xi_{{1}}}^{2}+{\xi_{{2}}}^
{2}\right )}}\  .
$$

{\bf 4.}\ The last example  illustrates Remark~I 
following the theorem (see Introduction).
In this case we consider the operator \eq{11} with the potential
\eq{13} associated with the proper $N$-soliton solution of the KdV
equation. We take  $ N=2 $ and fix $\, k_{1}=1,\, k_{2}=2\, $. 
The real phases are chosen as follows $\, \vartheta_{1}= 
\mbox{arctanh}\,(1/2) ,\, \vartheta_{2} =  \mbox{arctanh}\,(1/4)
\,$. The corresponding operator $ {\cal L}_{(k)} $ reads
$$
{\cal L}_{(k)} = 
\Box_{n+1} +
\frac {
 2 \left(2 x_{0} -3 x_{1}\right)
\left (3 x_{1}^{3}- 6 x_{0} x_{1}^{2}+
4 x_{1} x_{0}^{2}+8 x_{0}^{3} \right)
}
{ x_{1}^{2} \left(4 x_{0}^{2}-2 x_{0}x_{1}- x_{1}^{2}
\right)^{2}
}
\  .
$$
According to the theorem, it is huygensian provided
$ n $ is odd and $ n \geq 7 $. The nonzero Hadamard
coefficients are given by the following formulas:
$$
U_{{0}}=1,
$$
$$
U_{1}=
\frac {
4 \xi_{0}^{2}x_{1}^{2}+9 \xi_{1}^{2}x_{1}^{2}-
16 \xi_{0}^{2}x_{0}^{2}+8 \xi_{0}^{2} x_{0}x_{1}
}
{
2 x_{1}\,\left(
4 x_{0}^{2}-2 x_{0}x_{1}-  x_{1}^{2} \right)
\xi_{1}\left(4 \xi_{0}^{2}-2 \xi_{0} \xi_{1}- \xi_{1}^{2}
\right )
} +
$$
$$
+
\frac{
8 \xi_{0} \xi_{1} x_{0}^{2}-12 \xi_{1}^{2}x_{0} x_{1}
+ 4 \xi_{1}^{2} x_{0}^{2} -12 \xi_{0}\xi_{1}x_{1}^{2}
+16 \xi_{0}\xi_{1} x_{0} x_{1}
}
{
2 x_{1}\,\left(
4 x_{0}^{2}-2 x_{0}x_{1}-  x_{1}^{2} \right)
\xi_{1}\left(4 \xi_{0}^{2}-2 \xi_{0} \xi_{1}- \xi_{1}^{2}
\right )
}\ ,
$$

$$
U_{{2}}=-{\frac {
5\,\left (2\,\xi_{{0}}- \xi_{{1}}\right )\left (2\,x_{{0
}}-x_{{1}}\right )}{4\,x_{{1}}\left (4\,{x_{{0}}}^{2}-2
\,x_{{0}}x_{{1}}-{x_{{1}}}^{2}\right )\xi_{{1}}\left (4
\,{\xi_{{0}}}^{2}-2\,\xi_{{0}}\xi_{{1}}-{\xi_{{1}}}^{2}
\right )}}\ .
$$
\end{document}